\documentclass[%
 reprint,
 amsmath,amssymb,
 aps,
]{revtex4-2}

\usepackage{graphicx}
\usepackage{dcolumn}
\usepackage{bm}
\usepackage{CJKutf8}

\newcommand{\bp}{\mathrm{bp}}
\newcommand{\tot}{\mathrm{tot}}

\newcommand{\non}{\mathrm{nonPAM}}

\begin{document}
\begin{CJK*}{UTF8}{gbsn}
\title{Target search in the CRISPR/Cas9 system:\\ Facilitated diffusion with target cues}

\author{Qiao Lu (路桥)}
\author{Simone  Pigolotti}%
 \email{simone.pigolotti@oist.jp}
\affiliation{Biological Complexity Unit,\\ Okinawa Institute of Science and Technology Graduate University,\\ Onna, Okinawa 904-0495, Japan
}


\begin{abstract}
We study how Cas9, a central component of the CRISPR/Cas9 system, searches for a target sequence on the DNA. We propose a model that includes as key ingredients 3D diffusion, 1D sliding along the DNA, and the effect of short binding sequences preceding the target (protospacer adjacent sequences -- PAMs). This latter aspect constitutes the main difference with traditional facilitated diffusion of transcription factors.  We solve our model, obtaining an expression for the average search time of Cas9 for its target. We find that experimentally measured kinetic parameters are close to the values yielding an optimal search time. Our results rationalize the role of PAMs in guiding the search process, and show that Cas9 searches for its targets in a nearly optimal way. \end{abstract}

                      
\maketitle
\end{CJK*}

\section{\label{sec:level1}Introduction}

The CRISPR/Cas9 is an important system found in many bacterial species \cite{bonomo2018physicist}. It is commonly considered to be the bacterial immune system \cite{bonomo2018physicist,jiang2017crispr}, although this view has been recently questioned \cite{westra2020unclear}. The CRISPR/Cas9 became popular due to its potential for gene editing. Cas9 is a central protein of the CRISPR/Cas9 system. It is able to target a 20 base pair (bp) DNA target site that is complementary to a guide RNA (gRNA) that is loaded on it. After finding a matching DNA sequence, a hybrid forms between one strand of the target sequence and the gRNA, usually followed by cleavage of the target DNA sequence. Target recognition by Cas9 is necessarily triggered by a protospacer adjacent motif (PAM), that is a short DNA sequence preceding the target. For spCas9, a common variant of Cas9, the PAM sequence is "NGG", in which "N" can be any base. Such short motif is very abundant on genomes. For example, the {\em Escherichia coli} genome contains about half a million PAM sequences \cite{jones2017kinetics}. Cas9 binds on PAM sites in a tighter way than on other sites \cite{globyte2019crispr}. Moreover, experiments indicate that Cas9 can locate PAM sequences by sliding on the DNA \cite{globyte2019crispr}. 


Our focus is to theoretically understand how Cas9 searches for its target sequence. Previous models have focused on particular aspects of this dynamics. Farasat and Salis studied the process from PAM recognition to cleavage \cite{farasat2016biophysical}. Shvets and Kolomeisky proposed a model for the search dynamics of the target sequence, including binding/unbinding on PAM sequences but not sliding \cite{shvets2017mechanism}. Our previous work analyzed binding on PAM sequences by sliding, but without considering the target sequence \cite{lu2021search}.

A key aspect of the dynamics of Cas9 is the alternance of 3D diffusion and 1D sliding on the DNA. Such alternance has been widely studied in the biophysics literature, and it is commonly referred to as ``facilitated diffusion''. Facilitated diffusion was originally proposed  \cite{berg1981diffusion, winter1981diffusion, winter1981diffusion3} to explain how some transcription factors (TFs) are able to locate their target on the DNA faster than predicted by 3D diffusion alone \cite{riggs1970lac}. Indeed, the combination of 3D diffusion and sliding on the DNA can significantly speed up the search process \cite{berg1981diffusion, winter1981diffusion, winter1981diffusion3,mirny2009protein,sheinman2012classes}.  Energetics plays an important role in facilitated diffusion. Proper recognition requires a DNA-binding protein to bind its target site tightly, i.e., a large binding energy differences between the target and other binding sequences. However, such large energy differences result in a rough binding energy landscape, leading to a much less effective sliding \cite{zwanzig1988diffusion}. There is therefore a trade-off between speed and stability, which can be alleviated by different possible mechanisms  \cite{slutsky2004kinetics, mirny2009protein, sheinman2012classes, funnel}. 

\begin{figure}
\includegraphics[ scale = 0.12]{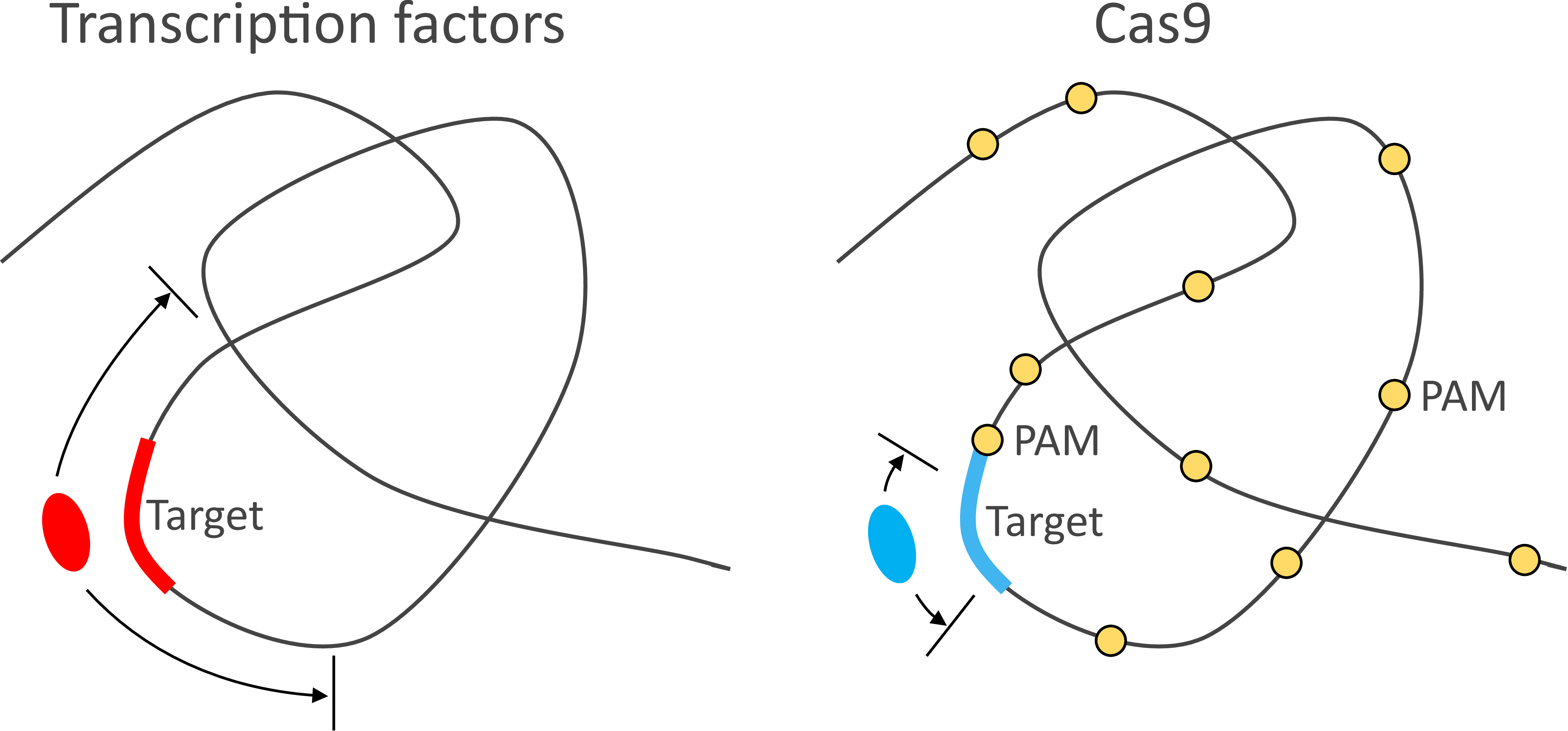}
\caption{\label{fig:compare}A comparison of TF and Cas9. Curved arrows indicate typical sliding distances in the two cases.}
\end{figure}

It is instructive to compare facilitated diffusion process of  TFs and Cas9, see Fig.~\ref{fig:compare}. Transcription factors usually have one or a few target sequence on the genome. The target of Cas9 is usually a single complementary sequence, but its recognition is triggered by one of the many PAM sequences on the genome. TFs can have a relatively long sliding length, usually estimated on the order of 100 bp \cite{winter1981diffusion3} or even more \cite{mirny2009protein}. In contrast, the sliding length of Cas9 is on the order of ten base pairs \cite{globyte2019crispr} or less \cite{lu2021search}. 

One could expect that PAM sequences could alleviate the limitation of a short sliding length, by providing some kind of guidance in finding the right target. This mechanism could be associated with
a speed-stability trade-off, as in the case of TFs. Binding too strongly to each PAM sequence would slow the searching task, but skimming too quickly over PAMs might cause a risk of missing the target sequence.

In this work, we theoretically study the average time it takes for Cas9 to find its target sequence on the DNA. Our model includes sliding, the role of PAMs, and realistic sliding energy landscapes. We analytically solve our model, obtaining an  expression for the average search time. We use this solution, combined with numerical simulations, to reveal the role of PAMs and sliding energetics on the efficiency of search. Our model reveals that, indeed, search in the presence of PAMs is characterized by a speed-stability tradeoff. We find that the optimal tradeoff is achieved when the PAM binding energy is of a few $k_BT$, in striking agreement with direct estimate of this binding energy. The length of the PAM sequence (2 bp) is also optimal for search. The optimal value of the detachment rate also well matches experimental measurements. Our results thus support that kinetic parameters characterizing Cas9 are poised close to the optimal performance for the search process.

\section{Model}\label{sec:model}

\begin{figure}
\includegraphics[ scale = 0.1]{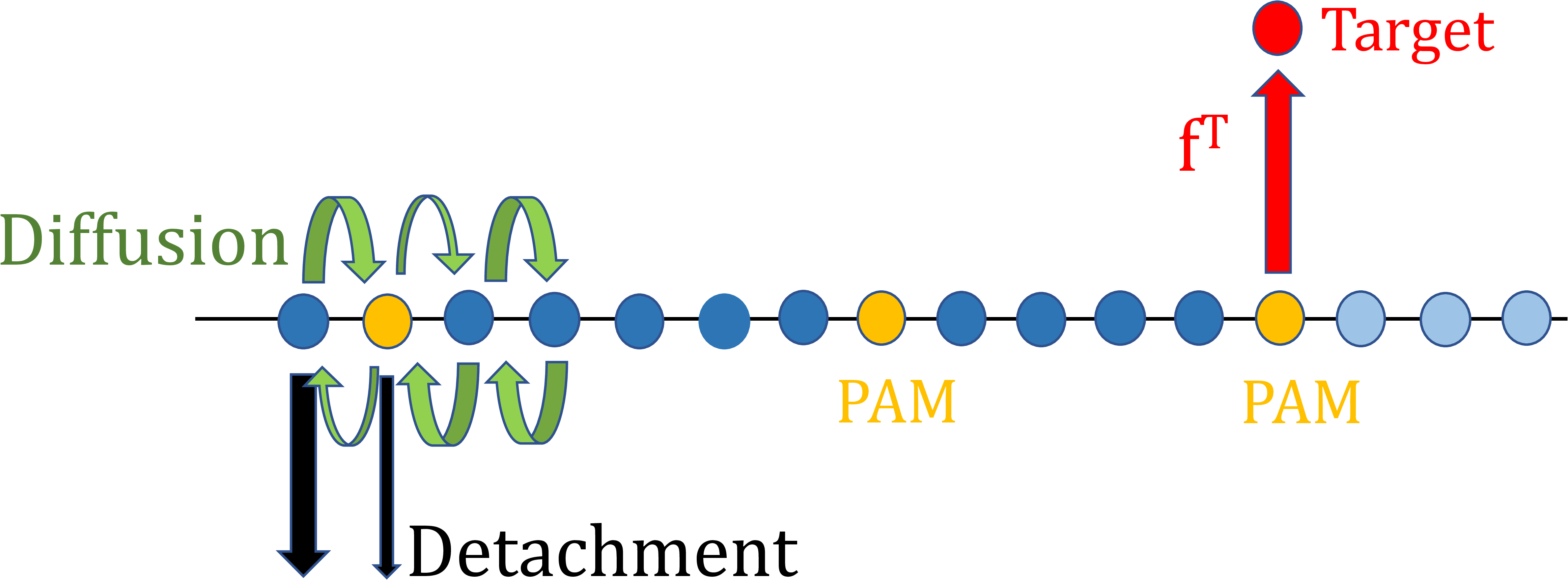}
\caption{\label{fig:sketch}A sketch of the model. Circles represent states of the Cas9. Blue and yellow circles represent base pairs on the DNA, and yellow circles represent the starting base pair of a PAM. Light blue circles represent the 20 bp target sequence.}
\end{figure}

We model the DNA as a one-dimensional lattice, in which each site represents a base pair (bp), see Fig~\ref{fig:sketch}. A Cas9 molecule can bind to one of the sites at a certain time. We distinguish two types of DNA sites: PAM sites are those correspond to the starting bp of the PAM sequence, while other bp, including the remaining bp of a PAM sequence, are considered as normal sites, see Fig~\ref{fig:sketch}. We call $E_n$ the binding energy of site $n$. Energies are expressed in units of $k_BT$, where $k_B$ is the Boltzmann constant and $T$ the temperature.

The Cas9 can detach or diffuse to either one of the neighbouring sites with rates
\begin{equation}
\begin{aligned}
k_n&=k\exp(E_n)\, ,\\
D_{m,n}&=D\exp(E_n)\qquad |m-n|=1,
\end{aligned}
\label{equation:chap4model}
\end{equation}
respectively. Here, $k$ and $D$ are rate constants. We consider a DNA of length $2L+1$ bp, such that $n\in [-L;L]$, with a unique target sequence complementary to the gRNA.  We assume that the PAM next to the unique target is located at the origin $n=0$. The boundary conditions are $D_{\pm(L+1),\pm L}=0$. The binding energy landscape $E_n$, of Cas9 on DNA is determined by the DNA sequence. For simplicity, we consider random DNA sequences, where each base pair occurs with probability $1/4$, except for the PAM located at the origin.

We assume that Cas9 can open a DNA bubble and form a hybrid between gRNA and a target sequence once bound to the central PAM. We call $f^T$ the rate at which this event occurs, where T stands for target (red arrow in Fig.~\ref{fig:sketch}). Once the target is reached, the search process is accomplished. 

If Cas9 detaches before finding the unique target, it performs a 3D diffusion round, with duration given by $t_{3D}$. This time is, in principle, a random quantity. After a 3D diffusion round, Cas9 can land on any DNA site with uniform probability and start its 1D search again. A search process always starts from a 1D diffusion with a uniformly distributed initial condition.

We determine the model parameters according to experimental evidence. In particular, we estimate $f^T$ from experimental results in \cite{3}. These results show that a hybrid forms approximately 100 ms after Cas9 binding, so we take $f^T=10\, s^{-1}$. 

We consider two kinds of energy landscape: 

\begin{itemize} 
\item {\bf Golf-course energy landscape,} In the golf-course energy landscape, non-PAM sites are characterized by a flat energy landscape, $E_n=0$. A PAM site has energy $E_n=\beta$, where $\beta$ is a parameter of the model.  We estimated the rate constants $k=1.94\, s^{-1}$, $D=52\, \bp^2 s^{-1}$, and the PAM energy $\beta=-3.34$ from single-molecule experiments \cite{globyte2019crispr}, see \cite{lu2021search}. 

\item {\bf Bp-dependent energy landscape.} In this case, all possible two bp sequences have different energy. This means that $E_n$ can take 16 possible values with equal probability. These values can be found in table~\ref{energy table}. For this energy landscape, the rate constants are $k=6.57\, s^{-1}$ and $D=160\, \bp^2 s^{-1}$, see \cite{lu2021search}. These rate constants are larger than in the golf course case because the baseline, i.e., the average energy of non-PAM sites, is different. Also in this case, we denote by $\beta$ the PAM binding energy.
\end{itemize}

\begin{table}[htb!]
\center
\begin{tabular}{p{1cm}|c|c|c|c}
& G & A & C & T\\ \hline
G & -4.47 & -2.61 & -1.12 & -1.42\\
\hline
A & -2.59 & -1.04 & -0.975 & -1.35\\
\hline
C & -1.22 & -1.40 & -1.35 & 0\\
\hline
T & -1.08 & -0.953 & -0.680 & -1.12\\
\end{tabular}
\caption{Cas9 binding energies $\Delta \epsilon_i$. Rows represent the first nucleotide and columns for the nucleotide next to the ``N''. The first element of the matrix represents the binding energy $\beta$ of a canonical PAM.  Details on how the energy values are derived from experimental evidences  \cite{boyle2017high} can be found in \cite{lu2021search}. }
\label{energy table}
\end{table}

In both cases, we take for $t_{3D}$ a constant value of $1.39 s$, since this value yields a roughly equal time of 1D and 3D diffusion for both kinds of landscape. This is theoretically considered to be the most efficient in facilitated diffusion \cite{slutsky2004kinetics,mirny2009protein,halford2004site,amir}.

\section{Average target search time}

In this Section, we estimate the average target search time in our model. We denote the (random) target search time by $T_\tot$ and by $\langle T_\tot \rangle$ its average. We use the notation $\langle\dots\rangle$ for an average over different realizations of the energy landscape, over trajectories, and over the initial position. We here present a relatively simple derivation of $\langle T_\tot \rangle$  that holds for large $L$. A more formal derivation can be found in Supplementary Information.

\subsection{Average time to reach the central PAM}

Our first step is to derive the average time $\langle T_0\rangle $ to reach the central PAM, without necessarily transitioning into recognition mode. This expression is an important ingredient to finally compute the average target search time. Our calculation makes use of a recent approach to facilitated diffusion based on conditional probabilities \cite{amir}, that we extend to a random energy landscape. 

We start by taking a continuous limit, and replace the discrete coordinate $n$ with a continuous coordinate $x$. We call $q(x)$ the probability to reach the origin within a 1D round before detaching, starting at coordinate $x$. Its expression is given by
\begin{equation}
\begin{split}
q(x)=\exp\left(-\sqrt{\frac{k}{D}}|x|\right)\, ,
\end{split}
\label{equation:pxsol4}
\end{equation}
as derived in \cite{amir}. In our model, this expression is independent of the distribution of $E(x)$. This fact is due to our choice of the diffusion and detachment rates in Eq.~\eqref{equation:chap4model}, such that disorder affects the timescales of diffusion and detachment, but not the probability of detaching or diffusing at a given site.

The average time to reach the central PAM is expressed by
\begin{equation}
\begin{split}
\left\langle T_0\right\rangle
=\frac{\left\langle q(x)t_{1D}(x)\right\rangle}{\left\langle q(x)\right\rangle}+\frac{\left\langle (1-q(x))t_{1D}^f(x)\right\rangle}{\left\langle q(x)\right\rangle}\\
+\left\langle t_{3D}\right\rangle \frac{1-\left\langle q(x)\right\rangle}{\left\langle q(x)\right\rangle}.
\end{split}
\label{equation:T0sol4}
\end{equation}
In this expression, the averages involving $q(x)$ are interpreted as averages over $x$, e.g., $\langle q(x)\rangle=\frac{1}{2L}\int^L_{-L}q(x)dx$. The quantities $t_{1D}(x)$ and $t_{1D}^f(x)$ are the durations of a 1D round that reached the origin or that failed reaching it before detaching, respectively. In facilitated diffusion, the average search time is often estimated as
\begin{equation}
\begin{split}
\langle T_0\rangle\approx \Gamma(\langle t_{1D}\rangle+\langle t_{3D}\rangle)\, ,
\end{split}
\label{equation:<T>funnel}
\end{equation}
see for example \cite{funnel}. Here, $\Gamma$ is the average number of 1D and 3D diffusion rounds before reaching the target. Comparing Eq.~\eqref{equation:T0sol4} with Eq.~\eqref{equation:<T>funnel}, $1/\langle q(x)\rangle$ plays the role of $\Gamma$; the first two terms in the right hand side represent the average time spent sliding; whereas the last term represents the average time spent performing 3D diffusion.

Evaluating Eq.~\eqref{equation:T0sol4}, we find that the average time $\langle T_0\rangle$ is expressed by
\begin{equation}
\begin{split}
\langle T_0\rangle =\overline{\exp(-E(x))}\frac{L}{\sqrt{kD}}+\langle t_{3D}\rangle L\sqrt{\frac{k}{D}},
\end{split}
\label{equation:<T0> result}
\end{equation}
in which the overline represents an average over binding sequences, weighted with their frequencies on the DNA. Eq.~\eqref{equation:<T0> result} is formally derived in Appendix~\ref{appa}. The first term in the right hand side of Eq.~\eqref{equation:<T0> result} represents the contribution from sliding and the last term the one from 3D diffusion. The factor $\overline{\exp(-E(x))}$ is the main difference between Eq.~\eqref{equation:<T0> result} and the main result of Ref.~\cite{amir}. This factor appears because, by Eq.~\eqref{equation:chap4model}, the characteristic dwell time on a position with energy $E(x)$ is proportional to $\exp(-E(x))$. Therefore, the total time spent sliding is proportional to $\overline{\exp(-E(x))}$, on average.

\subsection{Calculation of the total search time}

We now estimate the total search time $\langle T_\tot\rangle$. We make the approximation that, after Cas9 reaches the central PAM, it either finds the target, or diffuses to one of its neighbors. In other words, we neglect the possibility of directly detaching from the central PAM, since $ke^{\beta}\ll 2De^{\beta}\lesssim f^T$. As a consequence, a Cas9 at the central PAM finds the target, rather than diffusing away, with probability 
\begin{equation}
p=\frac{f^T}{f^T+2De^{\beta}}\, .
\label{eq:bindingapprox}
\end{equation}

We then define the average time to return to the origin $\langle T_1\rangle$, starting from one of the two nearest neighbouring sites of the origin. This can occur via a single 1D round, or after detachment and 3D diffusion rounds.

We now express the average total search time as a series:
\begin{equation}
\begin{split}
\langle T_\tot\rangle =& p\langle T_0\rangle+(1-p)p\left(\langle T_0\rangle+\langle T_1\rangle\right)\\
&+(1-p)^2p\left(\langle T_0\rangle+2\langle T_1\rangle\right)+...\\
=&\langle T_0\rangle+\frac{1-p}{p}\langle T_1\rangle.
\end{split}
\label{equation:<Ttot> series}
\end{equation}
To compute $\langle T_1\rangle$, we note that there are only two possibilities for a trajectory starting at $n=\pm 1$: reaching the origin before detaching, or detaching before reach the origin. We find that the probability of the former is $e^{-\sqrt{\frac{k}{D}}}$ and that of the latter is $(1-e^{-\sqrt{\frac{k}{D}}})$, see Appendix~\ref{appb}.

If Cas9 detaches before revisiting the origin, it would undergo a 3D diffusion round and spend again an average time $\langle T_0\rangle$ before reaching the origin. Therefore, we have
\begin{equation}
\begin{split}
\langle T_1\rangle = e^{-\sqrt{\frac{k}{D}}}\left(\langle t\rangle\right)+(1-e^{-\sqrt{\frac{k}{D}}})\left(\langle t'\rangle+\langle t_{3D}\rangle+\langle T_0\rangle\right).
\end{split}
\label{equation:<T1>}
\end{equation}
Here, $\langle t\rangle$ is the average time before revisiting the origin in the former case, and $\langle t'\rangle$ is the the average time until detaching in the latter case. We expect $\langle T_0\rangle$ to grow with $L$, whereas $\langle t\rangle$ and $\langle t' \rangle$ are on the order of the detachment time $k^{-1}$. Therefore, for large $L$, $\langle T_0\rangle$ is much larger than $\langle t\rangle$, $\langle t'\rangle$, and $\langle t_{3D}\rangle$. On the other hand, $e^{-\sqrt{\frac{k}{D}}}$ and $1-e^{-\sqrt{\frac{k}{D}}}$ are both of order 1. Therefore, we conclude that
\begin{align}
\langle T_1\rangle &\approx (1-e^{-\sqrt{\frac{k}{D}}})\langle T_0\rangle.
\label{equation:<T1>approx}
\end{align}
Substituting Eq.~\eqref{equation:<T1>approx} into Eq.~\eqref{equation:<Ttot> series}, we finally obtain
\begin{align}
  \langle T_\tot\rangle =& \left[1+\frac{1-p}{p}(1-e^{-\sqrt{\frac{k}{D}}})\right]\langle T_0\rangle\nonumber\\
  =&L\left[1+\frac{2De^{\beta}}{f^T}(1-e^{-\sqrt{\frac{k}{D}}})\right]\nonumber\\
  &\left(\frac{\overline{\exp(-E_n)}}{\sqrt{kD}}+\langle t_{3D}\rangle\sqrt{\frac{k}{D}}\right) ,
\label{equation:finals}
\end{align}
where for convenience we returned to the discrete notation $E_n$.

The more formal calculation presented in Supplementary Information leads to an expression for $\langle T_\tot\rangle$ that reduces to Eq.~\eqref{equation:finals} in the large $L$ limit. We found that, already for $L=2500$, the two solutions differ by less than $0.2\%$ for experimentally realistic parameters.

\section{Optimality of Cas9}

We now study how the mean total search time varies at changing parameters, and how the parameters leading to an optimal (i.e., minimum) total search time compare with the experimentally measured ones. 

\begin{figure}[hbt!]
\begin{center}	\includegraphics[ scale = 0.45]{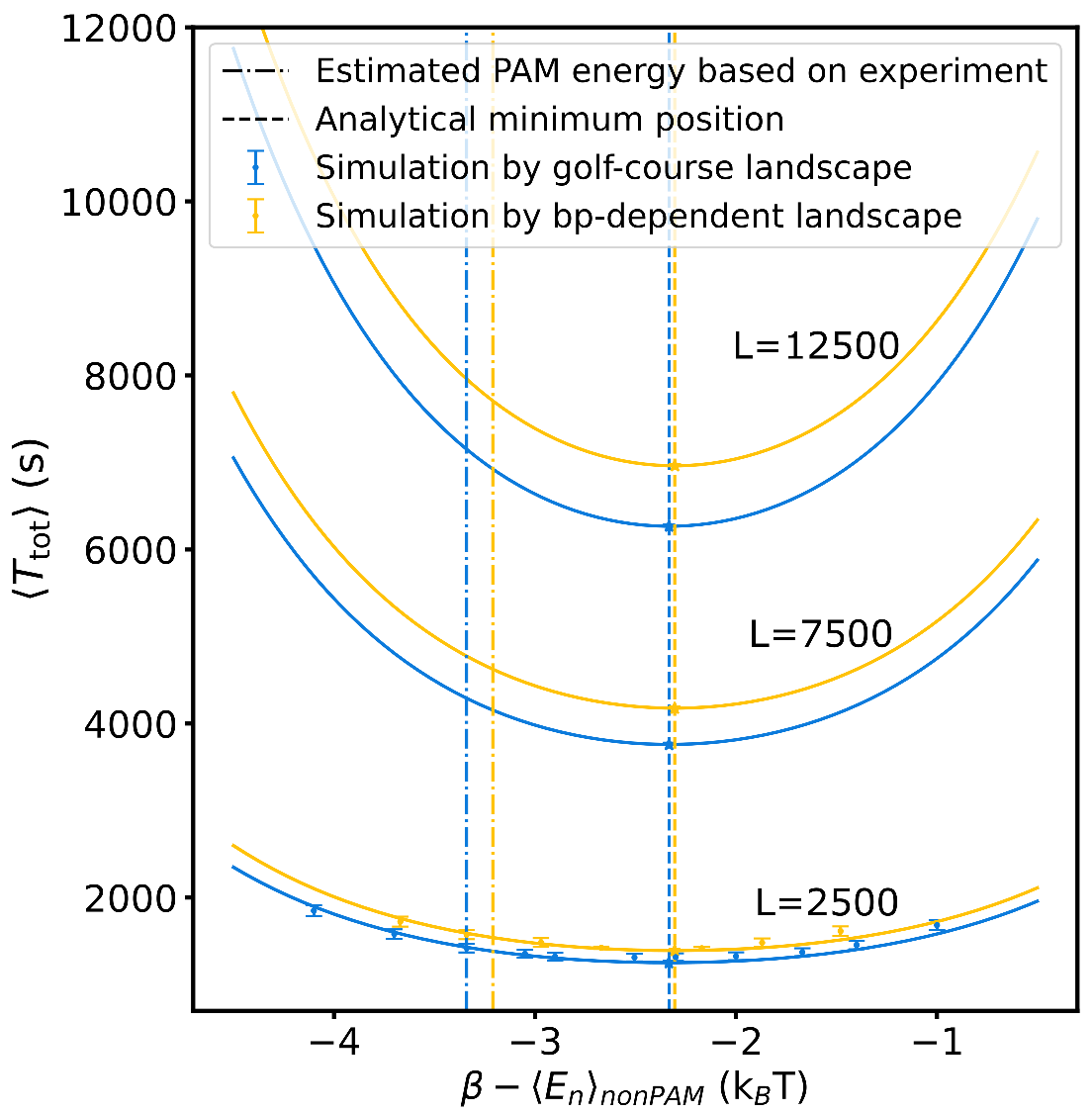}
\end{center}\vspace*{-20pt}\caption{Optimal PAM energy minimizes the total search time.  Blue color represents the golf-course landscape and yellow  the bp-dependent-kind. In the x axis, we plotted the difference between the PAM binding energy $\beta$ and the average binding energy $\langle E_n \rangle_\non$ of non-PAM sequences. The optimal PAM energy by theory is shown by stars and vertical dashed lines. The left two vertical lines are our estimation of the real PAM energy for both kinds of energy landscapes, see Ref.~\cite{lu2021search}. For $L=2500$, we compare the analytical result of Eq.~\eqref{equation:finals} with numerical simulations, finding an excellent agreement. }
\label{fig:check139}
\end{figure}

We start our analysis by varying the PAM binding energy $\beta$. We find that the average search time, as expressed by Eq.~\eqref{equation:finals}, presents a minimum as a function of $\beta$, see Fig.~\ref{fig:check139}.  This result supports our hypothesis of a speed-stability trade-off, i.e., the necessity of compromising between a sufficiently high binding energy (providing enough time to permit a transition to the recognition mode) and not spending too much time bound on the many PAM sequences.  Our model shows that this trade-off is very significant for large genomes. Since bacteriophage genome ranges from $5\times 10^3$ to $5\times 10^6$ bp \cite{mcgrath2007bacteriophage}, a typical invading genome can be even larger than those considered in Fig.~\ref{fig:check139}. In these cases, our model predicts that the average total search time $\langle T_\tot\rangle $ becomes very sensitive to the PAM energy.

By taking the derivative of Eq.~\eqref{equation:finals} with respect to $\beta$, we find the value $\beta^*$ of the PAM energy  that minimizes the total search time:
\begin{equation}
\begin{split}
\beta^*\approx\\
-\frac{1}{2}\ln\left[\frac{2D}{\alpha f^T}(1-e^{-\sqrt{\frac{k}{D}}})\left(\overline{\exp(-E_n)}-\alpha e^{\beta}  + t_{3D}k\right)\right],
\end{split}
\label{equation:betamin}
\end{equation}
in which $\alpha$ is the fraction of PAM sequences in the genome (e.g., $\alpha=1/16$ for a two-bp PAM in a random DNA sequence). We note that the quantity $(\overline{\exp(-E_n)}-\alpha e^{\beta})$ can be expressed as the average of $\exp(-E_n)$ over non-PAM sequences, and therefore does not depend on $\beta$. Since Eq.~\eqref{equation:finals} is linear in $L$, $\beta^*$ is independent of $L$, at least in the large $L$ limit. The value of $\beta^*$ predicted by Eq.~\eqref{equation:betamin} is very close to the experimentally determined value of $\beta$, see Fig.~\ref{fig:check139}. This supports that Cas9 search for its targets in a near-optimal way.

We now analyze the dependence of $\beta^*$ on model parameters. Varying one parameter on the right-hand side of Eq.~\eqref{equation:betamin} at one time, while keeping the others constant, we find the following relations. Firstly,  $\beta^*$ increases with $\alpha$. This is because the more abundant the PAMs are, the more time is wasted on PAMs that are not followed by the target. Increasing $\beta$ can compensate for this and hence leads to a smaller total search time. Secondly, $\beta^*$ decreases at increasing $k$ and $t_{3D}$. At larger $k$, the cost of missing the target is higher: the risk of detaching and doing another 3D and 1D round is larger. Decreasing $\beta$ alleviates this effect, therefore leading to a smaller $\langle T_\tot\rangle$. The reason is similar for $t_{3D}$: the cost of missing the target grows with the 3D diffusion time. Finally, since $D$ affects both speed and stability, the dependence of $\beta^*$ on $D$ is not monotonic.

We now study what would happen if the PAM sequence was made up of a number of base pairs other than two. For simplicity, we perform this analysis for the golf-course energy landscape only. When varying the number of base pairs in a PAM, we assume that $\beta$ is proportional to the number of base pairs, whereas other sites always have energy $E_n=0$. Both simulations and analytical result from Eq.~\eqref{equation:finals} show that a two-bp PAM leads to the smallest total search time, further supporting that the Cas9 binding mechanism is optimized to speed up search. If there were no PAM sequences at all, but Cas9 could still find recognize the target when on it,  the average search time would be twice as large compared to this optimal case.

\begin{figure}[hbt!]
\begin{center}
	\includegraphics[ scale = 0.5]{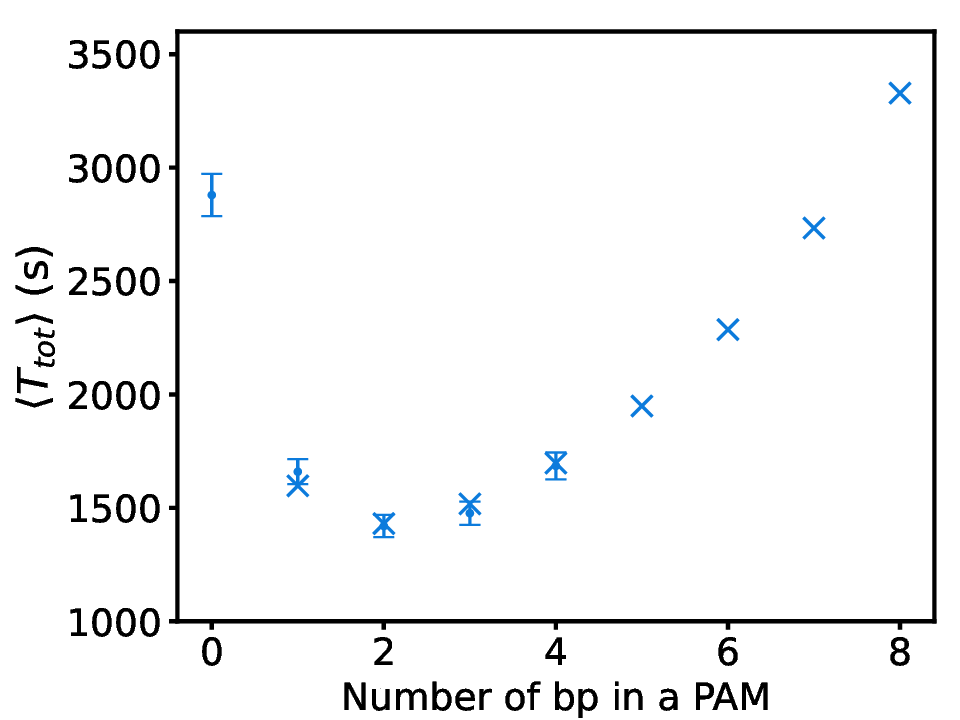}
\end{center}\vspace*{-20pt}\caption{Average total search time for hypothetical PAMs with different number of base pairs. Here, a two bp PAM has binding energy $\beta=-3.34$. The $x$ axis represents the number of bps in a PAM. The first point (zero bp) represents the case without PAM sequences guiding the search process. Points with error bars represent simulation results. Stars represent analytical predictions from Eq.~\eqref{equation:finals}.}
\label{fig:bp}
\end{figure}

\begin{figure}[hbt!]
\begin{center}
	\includegraphics[ scale = 0.5]{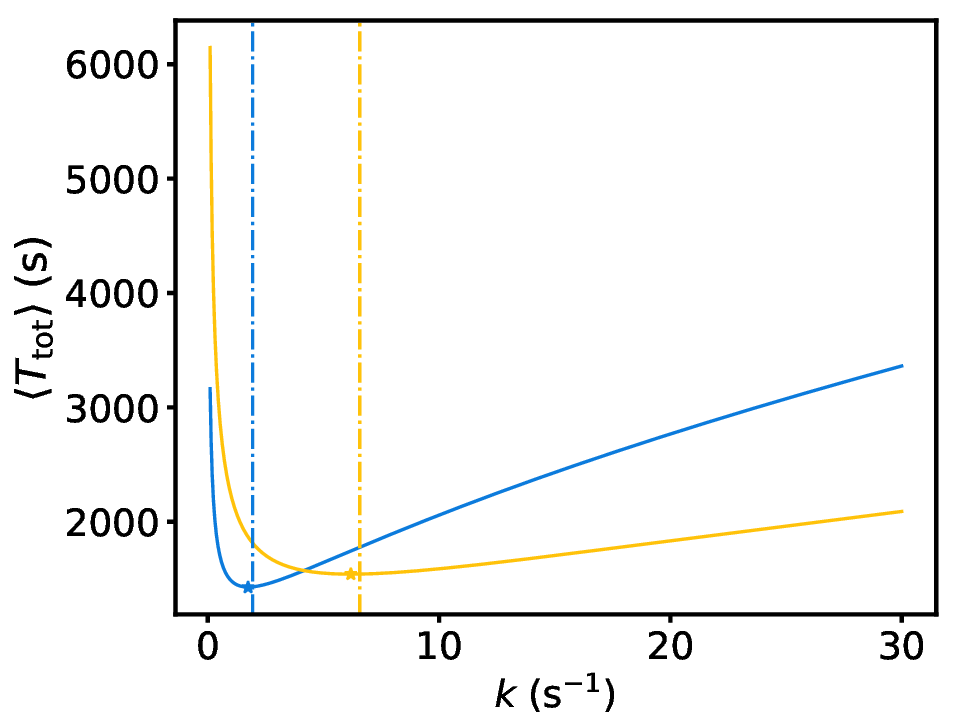}
\end{center}\vspace*{-20pt}\caption{Dependence of the average total search time on the unbinding rate.  
We fixed $L=2500$ since a larger $L$ simply scales the curve as in Fig.~\ref{fig:check139}. Color code is the same as in Fig.~\ref{fig:check139}. Stars mark the theoretical minimum. Vertical lines denote our estimate of the rate $k$ from experiments.}
\label{fig:k}
\end{figure}

Finally, we investigate the optimality of Cas9 with respect to the values of the unbinding rate $k$ and the 1D diffusion rate $D$. Our theory predicts a minimum in $\langle T_\tot\rangle$ at a value of $k$ that is very close to our estimation from experiments for both landscape kinds, see Fig.~\ref{fig:k}. 

\begin{figure}[hbt!]
\begin{center}
	\includegraphics[ scale = 0.5]{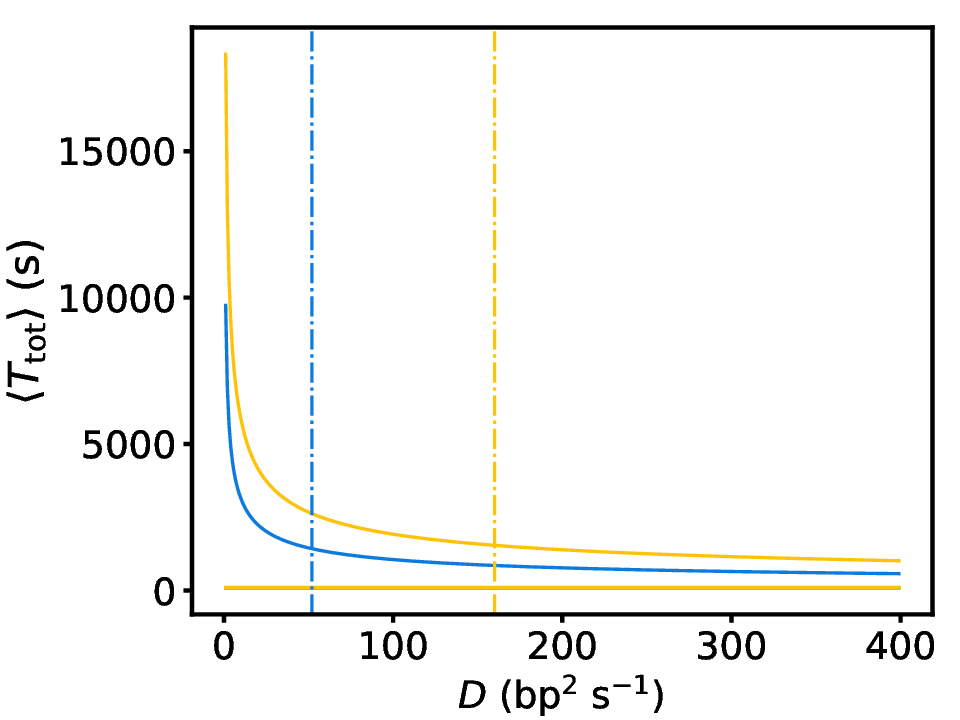}
\end{center}\vspace*{-20pt}\caption{Dependence of the total search time on the sliding rate $D$. The genome length is $2L+1$, with $L=2500$. Color code is the same as in \ref{fig:check139}. Horizontal lines mark the theoretical $\langle T_\tot\rangle$ values in the large $D$ limit (95.6 s for the golf-course, and 107.3 s for the bp-dependent landscape, respectively). Vertical lines mark the values of $D$ estimated from experiments, see \cite{lu2021search}.}
\label{fig:D}
\end{figure}

The total search time decreases monotonically with $D$. In the large $D$ limit, the total search time rapidly tends to an asymptotic value $\frac{2Le^{\beta}}{f^T}\left(\overline{\exp(-E_n)}+\langle t_{3D}\rangle k\right)$, see Fig.~\ref{fig:D}. This rapid convergence could explain why the sliding length of Cas9 is not very large when compared to that of TFs: increasing the sliding length by increasing $D$ might be evolutionary costly, without increasing performance very much. 

\section{Conclusion}

In this work, we theoretically studied how Cas9 searches for its target on the DNA. Our model quantifies the role of PAM sequencing in guiding the search process, and reveals trade-offs leading to optimal parameter values that are very close to experimentally measured ones. In particular, we have found that two base pair PAM sequences lead to the optimal search time. Experimentally inferred detachment rate and PAM binding energies are also remarkably close to the theoretically predicted optimal values.

An in-vivo experiment \cite{jones2017kinetics} found that an individual Cas9 in E.coli requires about 6 hours to find one out of 36 target sequences on the {\em E. coli} DNA. Our Eq.~\eqref{equation:finals} would predict 63.7 hours in the golf-course landscape, about one order of magnitude longer. In this estimate, we considered that the {\em E. coli} genome is 4.6 million bp long and contains about 0.5 million PAMs \cite{jones2017kinetics}. We have accounted for the 36 targets; for the fact that DNA is double-stranded (effectively doubling the genome length); and have used other parameter values given in Section~\ref{sec:model}. 

There are several possible explanations for this discrepancy. Our assumption that the total time spent by 1D diffusion is roughly the same as 3D diffusion may not hold. In fact the former is often found to be much larger than the latter, see, e.g., \cite{sheinman2012classes}. Other mechanisms not considered in our model, such as hopping
\cite{van2008dna,lomholt2009facilitated}, could also speed up the search process in vivo. Finally, our estimates of $f^T$, $k$ and $D$ are based on in vitro experiment, and these rate constants might substantially differ values in vivo. This is consistent with the observation that the inferred estimate of the PAM residence time in vivo from Ref.~\cite{jones2017kinetics} differs substantially from in vitro measurements. 

All of these causes could lead to an overestimation of the total search time, with the latter being probably the most important. It will be important in future work to clarify how the kinetic parameters of Cas9 differ between in vivo and vitro, and how the trade-offs revealed in our work affect the optimal parameters in these two cases.

\begin{acknowledgments}
We thank Zev Bryant for  discussions.
\end{acknowledgments}

\appendix

\section{Average time to reach the origin} \label{appa}

We calculate the average time $\langle T_0\rangle$ to reach the origin by facilitated diffusion, by extending the approach introduced in Ref.~\cite{amir}. For any quantity $a$, we denote the average over different landscape realizations by $\overline{a}$. This average may still be a function of the spatial coordinate. 

\subsection{Probability to find the target in a 1D round}

We call $q_n$ the probability to find the target in a 1D round, given that it started at site $n$. This quantity obeys a recursion relation 
\begin{equation}
\begin{split}
q_n=\frac{D_{n+1,n}}{D_{n+1,n}+D_{n-1,n}+k_n}q_{n+1}\\
+\frac{D_{n-1,n}}{D_{n+1,n}+D_{n-1,n}+k_n}q_{n-1}\, .
\end{split}
\label{equation:pn}
\end{equation}
By using the expression for $k_n$ and $D_{m,n}$ given in Eq.~\eqref{equation:chap4model} and rearranging terms,  Eq.~\eqref{equation:pn} becomes
\begin{equation}
\begin{split}
q_{n+1}+q_{n-1}-2q_n=\frac{k}{D}q_n
\end{split}
\label{equation:pnfinal}
\end{equation}
Since $p_0=1$ and assuming that the DNA chain is very long, the solution is
\begin{equation}
\begin{split}
q_n=\exp\left(-\alpha|n|\right)\, ,
\end{split}
\label{equation:pnsol}
\end{equation}
where $\alpha\equiv\cosh^{-1}(1+k/(2D))$. 

We now approximate the discrete DNA coordinate $n$ with a continuous coordinate $x$. In this approximation, we call $E(x)$,  $k(x)=k\exp(E(x))$, and $D(x)=D\exp(E(x))$ the binding energy, detaching rate, and diffusion rate at position $x$, respectively. The continuous counterpart of Eq.~\eqref{equation:pnfinal} reads
\begin{equation}
\begin{split}
\frac{d^2q(x)}{dx^2}=\frac{k}{D}q(x), 
\end{split}
\label{equation:pxfinal}
\end{equation}
where $q(x)$ is the continuous version of $q_n$. The solution of Eq.~\eqref{equation:pxfinal} reads
\begin{equation}
\begin{split}
q(x)=\exp\left(-\sqrt{\frac{k}{D}}|x|\right)\, .
\end{split}
\label{equation:pxsol}
\end{equation}
Equations~\eqref{equation:pnfinal} and \eqref{equation:pxfinal} are independent of the energy landscape, since the factors $\exp(E_n)$ or $\exp(E(x))$  cancel out. This means that $\overline{q}(x)=q(x)$ and $\overline{q}_n=q_n$.

\subsection{Mean first passage time and mean failed search time}

We denote by $T_n$ ($T(x)$) the mean first passage time (MFPT) to the origin, within a single 1D round and given that the Cas9 started at $n$ ($x$). We normalize this mean time by the total number of trajectories:
\begin{equation}
\begin{split}
T(x)=\lim_{\mathcal{N}\rightarrow\infty}\sum_{i=1}^{\mathcal{N}^s} \frac{t_{1D,i}(x)}{\mathcal{N}},
\end{split}
\label{equation:Ns}
\end{equation}
where $\mathcal{N}^s$ is the number of successful trajectories starting from $x$, $\mathcal{N}$ the total number of trajectories starting from $x$, and $t_{1D,i}(x)$ is the time spent by the $i$-th trajectory that reached $x=0$ before detach. Similarly, we denote by $T^f(x)$ the mean failed search time, given the protein started at $x$. This is the mean time of 1D diffusion contributed by trajectories that failed to reach the target before detachment, normalized by the number of all trajectories:
\begin{equation}
\begin{split}
T^f(x)=\lim_{\mathcal{N}\rightarrow\infty}\sum_{i=1}^{\mathcal{N}^f} t_{1D,i}^f(x)/\mathcal{N},
\end{split}
\label{equation:Nf}
\end{equation}
where $\mathcal{N}^f$ is the number of trajectories failed to reach the origin before detachment, and $t_{1D,i}^f(x)$ is the time spent by the $i$-th trajectory among them. By their definitions, $\mathcal{N}^s+\mathcal{N}^f=\mathcal{N}$. The discrete counterparts are similarly defined.

Both $T_n$ and $T^f_n$ are time-independent functions. The time $T_n$ obeys the recursion relation
\begin{equation}
\begin{split}
T_n=T_n(1-k_n\delta t-D_{n+1,n}\delta t-D_{n-1,n}\delta t)\\
+\delta tq_n+D_{n+1,n}\delta tT_{n+1}+D_{n-1,n}\delta tT_{n-1}
\end{split}
\label{equation:Tn}
\end{equation}
Using the expression for $k_n$ and $D_{m,n}$, Eq.~\eqref{equation:chap4model}, this simplifies to
\begin{equation}
\begin{split}
T_{n+1}+T_{n-1}-2T_n-\frac{k}{D}T_n+\frac{\exp(-E_n)}{D}q_n=0
\end{split}
\label{equation:Tnfinal}
\end{equation}
The continuous equivalent of Eq.~\eqref{equation:Tnfinal} is
\begin{equation}
\begin{split}
\frac{d^2T(x)}{dx^2}-\frac{k(x)}{D(x)}T(x)+\frac{1}{D(x)}q(x)=\\
\frac{d^2T(x)}{dx^2}-\frac{k}{D}T(x)+\frac{e^{-E(x)}}{D}q(x)=0.
\end{split}
\label{equation:TxfinalPAM}
\end{equation}
Imposing $E(x)=0$ except at the origin, Eq.~\eqref{equation:TxfinalPAM} becomes
\begin{equation}
\begin{split}
\frac{d^2T(x)}{dx^2}-\frac{k}{D}T(x)+\frac{1}{D}q(x)=0.
\end{split}
\label{equation:Txfinal}
\end{equation}
Given $q(x)$ as in \eqref{equation:pxsol}, and the boundary conditions $T(0)=0$, $T(\infty)=0$, Eq.~\eqref{equation:Txfinal} is solved by
\begin{equation}
\begin{split}
T(x)=\frac{|x|}{2\sqrt{kD}}\exp\left(-\sqrt{\frac{k}{D}}|x|\right).
\end{split}
\label{equation:Txsol}
\end{equation}

For general non-constant $E(x)$, we cannot directly solve Eq.~\eqref{equation:TxfinalPAM}. We can however compute $\langle T(x)\rangle$ as follows. The spatial average over $x$ and ensemble average commute, so we can first take the ensemble average of \eqref{equation:TxfinalPAM} to get $\overline{T}(x)$. We note that the ensemble average and derivatives commute, and since $\overline{q(x)}=q(x)$, $\overline{\exp(-E(x))q(x)}=\overline{\exp(-E(x))}q(x)$ we have
\begin{equation}
\begin{split}
\frac{d^2\overline{T}(x)}{dx^2}-\frac{k}{D}\overline{T}(x)+\frac{\overline{\exp(-E(x))}}{D}q(x)=0.
\end{split}
\label{equation:TxfinalPAMensemble}
\end{equation}
Since $\overline{\exp(-E(x))}$ is a constant independent of $x$ due our homogeneity assumption, Eq.~\eqref{equation:TxfinalPAMensemble} admits the same solution as  Eq.~\eqref{equation:Txfinal}, but with the average time multiplied by a factor $\overline{\exp(-E(x))}$:
\begin{equation}
\begin{split}
\overline{T}(x)=\overline{\exp(-E(x))}\frac{|x|}{2\sqrt{kD}}\exp\left(-\sqrt{\frac{k}{D}}|x|\right).
\end{split}
\label{equation:Txsolensemble}
\end{equation}

We now turn to the mean failed search time $T^f(x)$. This is the mean time of 1D diffusion contributed by those that failed to reach the target before detachment. The equation for $T^f(x)$ is the same as that of $T(x)$ but with $q(x)$ replaced by $1-q(x)$. The counterpart of \eqref{equation:TxfinalPAMensemble} is
\begin{equation}
\begin{split}
\frac{d^2\overline{T^f}(x)}{dx^2}-\frac{k}{D}\overline{T^f}(x)+\frac{\overline{\exp(-E(x))}}{D}(1-q(x))=0.
\end{split}
\label{equation:TfxfinalPAMensemble}
\end{equation}
and its solution reads
\begin{equation}
\begin{split}
\overline{T^f}(x)=\overline{\exp(-E(x))}\\
\left(-\frac{|x|}{2\sqrt{kD}}\exp\left(-\sqrt{\frac{k}{D}}|x|\right)+\frac{1}{k}\left[1-\exp\left(-\sqrt{\frac{k}{D}}|x|\right)\right]\right).
\end{split}
\label{equation:Tfxsolensemble}
\end{equation}

\subsection{Expression for $\langle T_{0}\rangle$}

We denote by $x_i$ the starting position of the i-th 1D diffusion round. We call $t_{1D}(x_i)$ the time spent in the successful i-th 1D search. We call $t_{1D}^f(x_i)$
the time spent in a failed i-th 1D search. Finally, we call $t_{3D}^i$ the time spent in the 3D search after the i-th failed 1D search. These three quantities are in principle stochastic.

We denote the first 1D diffusion round by subscript 0. Then the total search time $T_{0}$ can be expressed as
\begin{equation}
\begin{split}
T_{0}=q(x_0)t_{1D}(x_0)+(1-q(x_0))q(x_1)\times\\
\left(t_{1D}^f(x_0)+t_{1D}(x_1)+t_{3D}^0\right)\\
  +(1-q(x_0))(1-q(x_0))q(x_1)
  \times\\
  \left(t_{1D}^f(x_0)+t_{1D}^f(x_1)+t_{3D}^0+t_{3D}^1\right)+...\\
  =\sum^\infty_{i=0}q(x_i)\left(t_{1D}(x_i)+\sum^{i-1}_{j=0}\left(t^f_{1D}(x_j)+t_{3D}^j\right)\right)\times\\
  \prod^{i-1}_{j=0}\left(1-q(x_j)\right).
\end{split}
\label{equation:T0}
\end{equation}

We now take the average of Eq.~\eqref{equation:T0}. All 1D diffusion rounds begin from a uniformly distributed random position. Therefore, for any quantity $a$ in the expression \eqref{equation:T0}, we have $\left\langle a(x_i)\right\rangle=\left\langle a(x_j)\right\rangle$, so we drop the subscript and write them just as $\left<a(x)\right>$. 
By using that different 1D/3D rounds are independent, we obtain
\begin{equation}
\begin{split}
\left<T_0\right>=\sum^\infty_{i=0}\left( \left<q(x)t_{1D}(x)\right>\left(1-\left<q(x)\right>\right)^i\right.\\
\left.+i\left<q(x)\right>\left<t_{1D}^f(x)\left(1-q(x)\right)\right>\left(1-\left<q(x)\right>\right)^{i-1}\right.\\
\left.+i\left<q(x)\right>\left<t_{3D}\right>\left(1-\left<q(x)\right>\right)^i\right)\\
=\frac{\left<q(x)t_{1D}(x)\right>}{\left<q(x)\right>}+\frac{\left<(1-q(x))t_{1D}^f(x)\right>}{\left<q(x)\right>}+\left<t_{3D}\right>\frac{1-\left<q(x)\right>}{\left<q(x)\right>}.
\end{split}
\label{equation:T0sol}
\end{equation}

We evaluate $\langle q(x) \rangle$ using Eq.~\eqref{equation:pxsol}. The quantities $\left<q(x)t_{1D}(x)\right>$ and $\left<(1-q(x))t_{1D}^f(x)\right>$ are by definition equal to
$\frac{1}{2L}\int^L_{-L}\overline{T}(x)dx$ and $\frac{1}{2L}\int^L_{-L}\overline{T^f}(x)dx$, respectively. By using Eq.~\eqref{equation:Txsolensemble} and \eqref{equation:Tfxsolensemble} we obtain 
\begin{equation}
\begin{split}
\left<q(x)t_{1D}(x)\right>=\overline{\exp(-E(x))}\times\\
\left[-\frac{1}{2k}e^{-\sqrt{\frac{k}{D}}L}+\frac{1}{kL}\sqrt{\frac{k}{D}}(1-e^{-\sqrt{\frac{k}{D}}L})\right]
\end{split}
\end{equation}
and
\begin{equation}
\begin{split}
\left<(1-q(x))t_{1D}^f(x)\right>=\overline{\exp(-E(x))}\times\\
\left[\frac{1}{2k}e^{-\sqrt{\frac{k}{D}}L}-\frac{2}{kL}\sqrt{\frac{k}{D}}(1-e^{-\sqrt{\frac{k}{D}}L})+\frac{1}{k}\right].
\end{split}
\end{equation}
We substitute these three components into Eq.~\eqref{equation:T0sol} to find
\begin{equation}
\begin{split}
\langle T_0\rangle =&\overline{\exp(-E(x))}\left(\frac{L}{\sqrt{kD}(1-e^{-\sqrt{\frac{k}{D}}L})}-\frac{1}{k}\right)\\
&+\langle t_{3D}\rangle \left(\sqrt{\frac{k}{D}}\frac{L}{1-e^{-\sqrt{\frac{k}{D}}L}}-1\right)\\
\approx\,&\overline{\exp(-E(x))}\frac{L}{\sqrt{kD}}+\langle t_{3D}\rangle L\sqrt{\frac{k}{D}}.
\end{split}
\label{equation:<T0> result app}
\end{equation}
In this expression, we assumed that $L$ is large and neglected sublinear terms in $L$.

\section{Probability of reaching the origin before detaching} \label{appb}

We here derive the probability of reaching the origin before detaching, given the Cas9 started at $n=\pm1$. For simplicity, we perform this calculation in the continuous limit.

In a one-dimensional diffusion process, the first passage time distribution from any point $x_0$ to the origin is expressed by $\frac{x_0}{\sqrt{4\pi Dt^3}}e^{-\frac{x_0^2}{4Dt}}$. Simulation confirms that this continuous expression well approximates the discrete space case. We now compute the probability density $g(t)$ that the walker returns to $x=0$ at time $t$ before detaching. This density is obtained by setting $x_0=1$ in the first passage time distribution, and multiplying it by the probability $e^{-kt}$ that the walker has not detached yet:
\begin{equation}
\begin{split}
g(t)=\frac{1}{\sqrt{4\pi Dt^3}}e^{-\frac{1}{4Dt}-kt}.
\end{split}
\label{equation:g}
\end{equation}
It follows that our desired probability is
\begin{equation}
\begin{split}
\int^{+\infty}_0g(t)dt=\frac{1}{\sqrt{4\pi Dt^3}}e^{-\frac{1}{4Dt}-kt}dt=e^{-\sqrt{\frac{k}{D}}}.
\end{split}
\label{equation:magic}
\end{equation}
In this calculation, we have implicitly assumed the diffusion and detachment rates are constant in space. However, the results of Eq. \eqref{equation:magic} only depends on the ratio of $k$ and $D$, and by our model Eq.~\eqref{equation:chap4model}, this ratio is independent of $x$. Therefore, Eq.~\eqref{equation:magic} still hold for a non-constant energy landscape.


\bibliography{apssamp}

\end{document}